\documentclass[pra,twocolumn,showpacs]{revtex4}
\usepackage{epsfig}

\begin{document}

\title{Acceleration and localization of matter in a ring trap}

\author{Yu. V. Bludov$^1$ and V. V. Konotop$^{1,2}$}
\affiliation{$^1$Centro de F\'{\i}sica Te\'{o}rica e Computacional, Universidade de Lisboa,
 Complexo Interdisciplinar, Avenida Professor Gama Pinto 2, Lisboa
1649-003, Portugal
\\
  $^2$ Departamento de F\'{\i}sica, Faculdade de Ci\^encias,
Universidade de Lisboa, Campo Grande, Ed. C8, Piso 6, Lisboa 1749-016, Portugal and Departamento de
Matem\'aticas, E. T. S. de Ingenieros Industriales, Universidad de Castilla-La
Mancha 13071 Ciudad Real, Spain
}
%\date{ }

\begin{abstract}
A toroidal trap combined with external time-dependent electric field can be used for implementing different dynamical regimes of matter waves. In
particular, we show that  dynamical and stochastic  acceleration, localization and implementation  of the Kapitza pendulum can be originated  by
means of proper choice of the external force.

\end{abstract}

\maketitle

\section{Introduction}

Exploring different geometries of potentials trapping cold condensed atoms is of both fundamental and practical importance. Toroidal traps play a
special role allowing for "infinite" atomic trajectories and for realization of quasi-one-dimensional (quasi-1D) regimes. These advantages are
relevant for designing  highly precise sensors based on matter wave interferometry ~\cite{Gupta,Murch} as well as for accurate study of such
phenomena as superfluid currents, stability of sound waves,  solitons and vortices in Bose-Einstein condensates (BEC's)~\cite{SPR,Brand}. Traps with
circular geometry are also believed to be conceptually important for implementation of the main ideas of the accelerator physics at ultra-low
temperatures~\cite{Murch} and, in particular, for acceleration of ultracold atoms~\cite{Dutta}. In this last context existence of well localized wave
packets, and thus attenuation of the dispersion, the latter being the intrinsic property of a quantum systems, is of primary importance. In the first
experimental studies~\cite{Murch} it was shown that the dispersive spreading out~\cite{Gupta} can be compensated by using betatron resonances in a
storage ring. An alternative way of contra-balancing dispersion is also well known - it is nonlinearity, leading in quasi-1D regime to existence of
bright  and dark matter solitons (see e.g.~\cite{multiple1,bright} and \cite{Tsuzuki,dark,multiple2}, respectively). This issue has already been
explored~\cite{sol_ac} from the point of view of  acceleration of matter waves in a toroidal trap with help of a modulated optical lattice, which is
known to be an efficient tool for acceleration of matter waves~\cite{BKK}.

In this paper we propose two alternative ways of accelerating matter wave solitons -- either by time varying or by stochastic external electric
field. These new ways of soliton acceleration are especially relevant in view of radiative losses~\cite{Skryabin} and distortions~\cite{BKK} of
solitons moving in optical lattices (the effects acquiring significance for long trajectories). At the same time, it turns out that the toroidal
geometry of a trap confining a BEC allows one to realize a number of other dynamical regimes, like dynamical localization of solitons and solitonic
implementation of the celebrated Kapitza pendulum. Theoretical description of all mentioned phenomena can be put witching the unique framework, based
on the perturbation theory for solitons, what is done in the present paper. More specifically, in Sec.~\ref{sec:scaling} we formulate the model and
the main physical constraints determining its validity. In sections ~\ref{sec:dyn_accel} and~\ref{sec:stoch_accel} we describe how by applying
external time-dependent electric field matter solitons can be accelerated in the usual sense and in the sense of the time increase of the velocity
variance (the stochastic acceleration), respectively. In Sec.~\ref{sec:dyn_loc} we describe localized states of the matter in circular trap subject
to external field, and in Sec.~\ref{kapitza} we show that a matter soliton affected by rapidly varying  force represents an example of the Kapitza
pendulum~\cite{LL}. Summary and discussion of the results are given in the Conclusion.

\section{Scaling and the evolution equation}
\label{sec:scaling}

We assume that a BEC is loaded in a circular trap, which in cylindrical coordinates ${\bf r}=(\rho,\varphi,z)$ is described by
$V=V_{c}(\rho)+m\omega_z^2z^2/2$, where $\omega_z$ is the frequency of the magnetic trap in the $z$--direction, $V_{c}(\rho)$ is the potential in the
radial direction, forming the trap circular in the $(x,y)$--plane, and $m$ is the mass of an atom. We also suppose that the BEC is  subject to
external electric field with amplitude $E_0$, which produces an additional potential $V_{ext}=-\alpha^\prime E_0^2/4$, where $\alpha^\prime$ is the
polarizability of the atoms (see e.g.~\cite{Pet_Sm}). If the amplitude $E_0$ or direction of the field vary along some direction, say, along the
$x$-axis, smoothly on the scale of the trap radius $R$, the potential energy $V_{ext}$ van be expanded in  the Taylor series and, after neglecting
nonessential constant, be rewritten in the form $V_{ext}=-\alpha x$, where $\alpha=(\alpha^\prime /4)\partial (E_0)^2/\partial x|_{x=0}$ and we
restricted the consideration only to the first term of the expansion. In order to realize one-dimensional geometry we  require torus radius to be
much larger than the core radius $r_c$, what allows us to define a small parameter $\epsilon=r_c/R\ll 1$. In order to introduce quantitative
characteristics, we consider the normalized ground state $\phi$ of the eigenvalue problem
\begin{eqnarray}
\label{linear1}
-\frac{\hbar^2}{2m}\frac{1}{\rho}\frac{d}{d\rho}\rho \frac{d}{d\rho}\phi+V_c(\rho)\phi =\varepsilon_r\phi\,,\quad \int_0^\infty \phi^2 \rho d\rho=1
\end{eqnarray}
and define $R_1=\int_{0}^{\infty}\phi^2\rho^2 d\rho$, $R_2= \left(\int_{0}^{\infty}\phi^2d\rho/\rho\right)^{-1/2}$, and
$\lambda=\left(\int_{0}^{\infty}\phi^4\rho d\rho\right)^{-1/2}$. In the case at hand $\lambda \sim \sqrt{R r_c}\sim \epsilon^{1/2}R$ and thus
$\lambda\ll R_1\sim R_2\sim R$.

In the present paper we are interested in the dynamics of matter waves which spatial extension is much less than the trap perimeter, allowing treat them similarly to the matter solitons in an infinite one-dimensional trap. This in particular the case where the spatial size of the BEC excitations along the trap are of order of $\lambda$, which is the well defined parameter and thus convenient for formulation the constraints of the theory. Indeed,  now we can estimate  the kinetic energy of the longitudinal excitations as $\varepsilon_\|=\hbar^2/(2m\lambda^2)$ and require it to be much less than the kinetic energy of the transverse excitations, $\varepsilon_r\sim\hbar^2/(2mr_c^2)$ (for the sake of simplicity here we assume that the size of the trap in $z$-direction is of order of the core radius: $a_z=\sqrt{\hbar/m\omega_z}\sim r_c$).  Adding the requirement for the
energy of the two-body
interactions, which is estimated as $|g|n$ (where $g=4\pi \hbar^2 a_s/m$, $a_s$ is the scattering length, $n\sim N/{\cal V}$ is a mean density,
 $N$ is the total number of atoms and ${\cal V}$ is the effective volume occupied by the atoms and estimated as ${\cal V}\sim \pi\lambda r_ca_z$), to be of order of $\varepsilon_\|$ and to be much less than $\varepsilon_r$ (or more precisely, requiring
$|g|n/\varepsilon_r\sim \epsilon$), we can neglect in the leading order the transitions between the transverse energy levels~\cite{dark,multiple2}, and employ the multiple scale expansion~\cite{multiple1,multiple2} for description of the quasi-one-dimensional evolution of the BEC.
We also notice that subject to the assumptions introduced, one has the estimate $N\sim\epsilon^{3/2}R/(8|a_s|)$.

In order to get an insight on practical numbers, let us consider $^7$Li atoms ($a_s=-2\,$nm) in a trap with $R=100\,\mu$m, $r_c=5\,\mu$m and $a_z=10\,\mu$m. Then   $\epsilon=0.05$, the characteristic size of solitonic excitations is $\lambda\approx
22\,\mu$m and the number of particles is estimated as $N\approx 140$. We emphasize, that these estimates indicate only an order of the parameters. Thus, for
example, a condensate of $10^2\div 10^3$ lithium atoms satisfy the conditions of the theory.

We will be interested in managing soliton dynamics by means of weak (i.e not destroying solitons) electromagnetic field varying in time. Respectively, we consider  $\alpha$ time-dependent and characterized by the estimate $\alpha\sim \hbar^2/(mR_1\lambda^2)$. Then, starting with the Gross-Pitaevskii equation, in which the external potential in cylindrical coordinates has the form $V_{ext}=-\alpha\rho\cos(\varphi)\approx -\alpha R_1\cos(\varphi)$,  and using the multiple-scale
expansion one ensures that the BEC macroscopic wave function in the leading order allows factorization
\begin{eqnarray}
\Psi= \pi^{-1/4}a_z^{-1/2}e^{-i(\omega_r +\omega_z)t/2}e^{-z^2/2a_z^2} \phi(r)\psi(t,\varphi),
\end{eqnarray}
where $\omega_r=2\varepsilon_r/\hbar$ and $\psi(t,\varphi) $ solves the nonlinear Schr\"odinger equation, which we write in terms of
%\begin{eqnarray}
$ A=\sqrt{|g| m/\sqrt{2\pi}\hbar^2 a_z}\psi, $
%\end{eqnarray}
$\zeta= R_2\varphi/\lambda$, and $\tau=\hbar t/m\lambda^{2}$
\begin{eqnarray}
    \label{NLS}
    i \frac{\partial A}{\partial \tau}=-\frac{1}{2} \frac{\partial^2 A}{\partial \zeta^2}-\cos(\kappa\zeta)f(\tau) A+\sigma |A|^2A\,.
\end{eqnarray}
Here $\sigma=$sign$a_s$, $f(\tau)\equiv  mR_1\lambda^{2}\alpha (t)/\hbar^2$ and $\kappa=\lambda/R_2 \sim \sqrt{\epsilon}$. We choose the scaling  in
such a way that all terms in (\ref{NLS}) are of the unity order, and in particular $A={\cal O}(1)$. This can be done, taken into account the normalization
\begin{eqnarray}
\int_0^L |A|^2d\zeta=2\sqrt{2 \pi}\frac{|a_s|N}{\kappa a_z},
\end{eqnarray}
$L=2\pi/\kappa$,
 which follows from the normalization condition for the order parameter
$\int|\Psi|^2d^3{\bf r}=N$,   and considering $N\sim a_z/|a_s|$, what is of order of $10^{3}$, in a typical experimental setting.

Eq. (\ref{NLS}) is subject to periodic boundary conditions $A(\zeta,\tau)=A(\zeta+L,\tau)$.

\section{Acceleration of bright matter solitons by time-dependent external force}
\label{sec:dyn_accel}

First we consider the acceleration, $\gamma$, which can be achieved due to the potential $V_e$ properly dependent on time. An order of magnitude of
$\gamma$ can be estimated by taking into account that Eq. (\ref{NLS}) makes sense provided that all terms are of the unity order. In the physical
units this gives $\gamma\sim\hbar^2/(m^2\lambda^3)$. Then, recalling the above example of the lithium condensate we estimates $\gamma \sim
7\,$mm/s$^2$, what is of order of the acceleration  announced in~\cite{sol_ac}. It  however does not provide the best estimate in our case, because
it is based on the 1D model, while lowering dimensionality imposes constrains on the atomic density and consequently on the amplitude of the applied
force.

To describe the physics of the phenomenon we consider a BEC with a negative scattering length ($\sigma=-1$). Then a "bright soliton" solution of Eq.
(\ref{NLS}) at $f(\tau)\equiv 0$ (or more precisely a periodic solution mimicking a bright soliton in an infinite 1D system) which moves with a
constant velocity $v_n$ can be written down as follows~\cite{Tsuzuki}
\begin{eqnarray}
\label{bs}
    A_s=e^{-i(\omega(k)+v_n^2/2)\tau+iv_n\zeta}\eta(k)\,\mbox{dn}\left(\eta(k)(\zeta-v_n\tau) ,k\right).
\end{eqnarray}
Here $\mbox{dn}(x,k)$ is the Jacobi elliptic function~\cite{AS}, $k$ is the elliptic modulus parameterizing the solution. The frequency and the
amplitude are given by:   $\omega(k)=(k^2/2-1)\eta^2(k)$ and $\eta(k)=2 K(k)/L$ [$K(k)$ is the complete elliptic integral of the first kind].  The
velocity of the soliton is quantized  $v_n=2\pi n/L$ with $n$ being integer.

To ensure that the solution $A_s$ satisfies the scaling relations imposed above, we notice that the size of the soliton can be estimated as
$\pi/K(k)$ and its smallness implies that $k$ is close to unity. In that case we obtain the estimates $1-k^2\sim 16\exp(-2\pi/\sqrt{\epsilon})$ and
$\mbox{dn}\left(\eta(k)(\zeta-v_n\tau),k\right)\approx 1/\cosh (\eta(k)(\zeta-v_n\tau))$.

In the limit $k\to 1$ quantization of the velocity does not play any significant role. We verified this numerically. For example, for $L=10$
deviation of the initial velocity from the quantized one  produces appreciable  effect on dynamics during intervals $\tau\lesssim 100$ only if
$k\lesssim 0.99$.

When external force is applied, $f(\tau)\neq 0$, the velocity is not preserved any more, what manifests itself in evolution of the momentum $
P=(1/2i)\int_{0}^L \left(A_\zeta\bar{A}-A \bar{A}_\zeta\right)$ (here $\bar{A}$ stands for complex conjugation of $A$) according to the law:
\begin{eqnarray}
\label{p}
    \frac{dP}{d\tau}=-f(\tau)\int_0^L\cos(\kappa\zeta)\frac{\partial |A|^2}{\partial\zeta}d\zeta\,.
\end{eqnarray}
The external field, however, does not affect the norm:   ${\cal N}=\int_0^L|A|^2d\zeta$=const.
It follows from (\ref{bs}) that in the adiabatic approximation the solution of the perturbed equation (\ref{NLS}) can be searched in the form
 \begin{eqnarray}
\label{bs_ad}
    A=e^{-i(\omega(k)+V(\tau)^2/2)\tau+iV(\tau)\zeta}\eta\,\mbox{dn}\left(\eta(\zeta-X(\tau)), k\right)
\end{eqnarray}
where $V(\tau)=dX(\tau)/d\tau$ is the time-dependent velocity of the soliton and $X(\tau)$ is the coordinate of the soliton center. Substituting
(\ref{bs_ad}) in (\ref{p}) and taking into account the parity of the functions in the integrand as well as the fact that all of them are periodic
with the same period $L$, we obtain the  equation for the soliton coordinate
%of the soliton center
\begin{eqnarray}
\label{x}
    \frac{d^2X}{d\tau^2}= -\kappa C(k)f(\tau)\sin(\kappa X) \,. \label{eq:non-osc}
\end{eqnarray}
Here
\begin{eqnarray}
    C(k)= \frac{K(k)}{2\pi E(k)}\int_{0}^{2\pi}\cos(\theta)\mbox{dn}^2\left(\frac{K(k)}{\pi}\theta,k\right)d\theta
\end{eqnarray}
and it is taken into account that
%for solution (\ref{bs_ad})
${\cal N}=2\eta E(k)$, where $E(k)$ is the complete elliptic integral of the second
kind.

Depending on the choice of the function $f(\tau)$, Eq. (\ref{x}) describes different dynamical regimes. Now we are interested in acceleration which
occurs during the rotational movement of the soliton in the trap (i.e. $X$ is a growing function). We illustrate this acceleration using an example
of the simplest step-like dependence $f(\tau)$. To this end we assume that initially the soliton is centered at $X(0)=0$ and require $f(\tau)$ to be
a constant $f_0$ for time intervals such that the soliton coordinates $X(\tau)\in I_p$ and to be zero for $X(\tau)\notin I_p$ where the intervals
$I_p$ are given by $I_p=[(p+\frac 12)L,(p+1)L]$ with $p=0,1,...$. Then, as it is clear, the acceleration of the soliton, which is given by the right
hand side of (\ref{x}), is positive for all times. The above requirement introduces natural splitting of the temporal axis in the set of intervals
$T_l=[\tau_l,\tau_{l+1}]$ ($l=0,1,...$), with $\tau_0=0$, such that $f(\tau)=0$ for $\tau\in T_{2p}$ and $f(\tau)=f_0$ for $\tau\in T_{2p+1}$ (here
$X(\tau_l)=lL/2$). Thus, our task is to find $\tau_l$. This  can be done by taking into account that during each of the "odd" intervals $T_{2p+1}$
Eq. (\ref{x}) describes conservative nonlinear oscillator, the solution for which is well known. During "even" intervals $T_{2p}$ the motion is free
(with a constant velocity) what means that  the time $T_{2p}$ necessary for the soliton to cross an interval $[pL,(p+\frac 12)L]$ is
\begin{eqnarray}
T_{2p}=\tau_{2p+1}-\tau_{2p}=L/(2v_{2p}),\label{eq:2p}
\end{eqnarray}
where $v_{2p}$ is the velocity in the point $pL$. During the time interval $T_{2p+1}$ the soliton has to cross the interval $I_{p}$. From this
condition we obtain:
\begin{equation} T_{2p+1}=\tau_{2p+2}-\tau_{2p+1}=\frac{
\sqrt{2}K\left(\sqrt{2E_0/(H_{2p+1}+E_0)}\right)}{\kappa\sqrt{H_{2p+1}+E_0}},\label{eq:2pp1}
\end{equation}
where $H_{2p+1}=v_{2p+1}^2/2+E_0$ is the energy of the soliton in the point $(p+1/2)L$, $E_0=C(k)f_0$, and $v_{2p+1}=v_{2p}$ is the soliton velocity
in the same point. At the end of the interval $T_{2p+1}$ the soliton velocity is given by $v_{2(p+1)}=\sqrt{2(H_{2p+1}+E_0)}$. Thus one computes that
after $p$ rotations the soliton acquires the velocity $v_{2(p+1)}$, which can be obtained from the recurrent relation:
$v_{2(p+1)}=\sqrt{v_{2p}^2+4C(k)f_0}$.

\begin{figure}
  \begin{center}
   \begin{tabular}{c}
        %\scalebox{1.0}[1.0]{\includegraphics[28,586][269,700]{fig1.ps}}
        \scalebox{1.0}[1.0]{\includegraphics{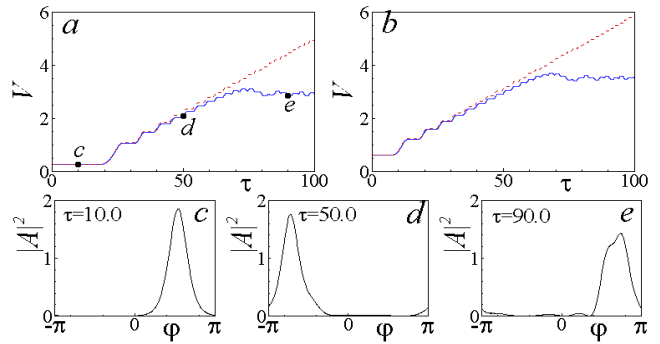}}
   \end{tabular}
   \end{center}
\caption{(Color online) The soliton velocity {\it vs} time (panels a,b) for the parameters $k=0.99999$, $L=10.0$, $f_0=0.3$, $n=0.43$ (panel a, the
non-quantized velocity), $n=1.0$ (panel b, the quantized velocity), and the forms of the soliton (panels c,d,e) in the instants of time indicated in
the panel a. In the panels a,b solid and dashed lines represent the  velocity numerically computed from Eq.(\ref{NLS}), and   Eq.(\ref{eq:non-osc}),
respectively. } \label{fig:accel-form}
\end{figure}

In Fig.\ref{fig:accel-form} a,b we compare the solution, obtained from the perturbation theory, Eq.(\ref{eq:non-osc}), with numerical simulation of
Eq.(\ref{NLS}) for $f_0=0.3$. Nevertheless during the numerical simulation we used the values for $T_{2p}$ and $T_{2p+1}$
(Eqs.(\ref{eq:2p})--(\ref{eq:2pp1})) obtained for the case of adiabatic approximation. It follows form the results presented that the dashed and
solid lines perfectly match until $\tau \approx 50.0$. At larger times appreciable discrepancy appears. It occurs due to failure of the adiabatic approximation and can be removed by introducing temporal corrections to $T_{2p}$ and $T_{2p+1}$. This naturally leads to an optimization problem, which requires numerical approach and goes beyond the scope of the present work. Finally we notice, that for the above example of $^7$Li condensate the obtained acceleration is $0.36$mm/s$^2$.

Comparison of the panels a and b in Fig.\ref{fig:accel-form} shows that for $k\approx 1$ quantization of the velocity is not important, what is also confirmed by the evolution of the solitonic forms depicted in the panels
Fig.\ref{fig:accel-form} c-e.

\section{Stochastic acceleration of matter solitons}
\label{sec:stoch_accel}

Now we concentrate on another dynamical regime -- on the {\em stochastic acceleration} -- where increase of the velocity of a matter soliton in a
toroidal trap is achieved by applying a fluctuating external field. To this end holding all conditions of the applicability of the model (\ref{NLS}),
we consider the case of a stochastic force $f(\tau)$, which is delta-correlated Gaussian process with characteristics $\langle f(\tau)\rangle=0$ and
$\langle f(\tau)f(\tau')\rangle=D\delta(\tau-\tau')$ (the angular brackets stand for the stochastic averaging and $D$ is the dispersion). Now the
dynamics can be described in terms of the distribution function
\begin{eqnarray}
 {\cal P}(V,\Phi,\tau)=\langle\delta(\Phi-\Phi(\tau))\delta(V-V(\tau))\rangle,
\end{eqnarray}
  where
$\Phi(\tau)\equiv \kappa X$ is the angular coordinate of the soliton, $\Phi(\tau)$ and $V(\tau)$ with explicit time dependence stand for the soliton
coordinates obtained from the dynamical equations while $\Phi$ and $V$ are considered as independent variables. The distribution function solves the
Fokker-Planck equation, which is obtained by the standard procedure (see e.g.~\cite{KV}):
\begin{eqnarray}
\label{FP}
    \frac{\partial {\cal P}}{\partial \tau}=-V\frac{\partial {\cal P}}{\partial \Phi}+\tilde{D}\sin^2(\Phi)\frac{\partial^2{\cal P}}{\partial V^2}\,.
\end{eqnarray}
Here $\tilde{D}=\kappa^4 C^2(k)D$ is the diffusion coefficient. Due to the circular geometry of the trap Eq. (\ref{FP}) is considered on the interval
$-\pi<\Phi<\pi$ with the periodic boundary conditions ${\cal P}(V,\Phi-\pi,\tau)={\cal P}(V,\Phi+\pi,\tau)$ with respect to $\Phi$ and zero boundary
conditions with respect $V$: ${\cal P}\to 0$ as $V\to\pm\infty$.  The normalization condition for the probability density reads:
$\int_{-\infty}^{\infty}dV\int_{-\pi}^{\pi}d\Phi {\cal P}=1$.

Multiplying  Eq.(\ref{FP}) by $V$  and $\Phi$ and integrating over $V$ and $\Phi$  one readily obtains that the averaged velocity and angular
position of the soliton are constants, which for the sake of simplicity will be considered zeros, i.e.  $\langle V\rangle=0$ and  $\langle
\Phi\rangle=0$. Next, multiplying (\ref{FP}) by $V^2$, $\Phi^2$ and $V\Phi$ and performing the integration one obtains the equations of the second
momenta. They are not closed and can be written down as follows:
\begin{eqnarray}
\label{V1}
&&\frac{d}{d\tau}\langle V^2\rangle =2\tilde{D}\langle \sin^2\Phi\rangle\,,
\\
\label{P1}
&&\frac{d}{d\tau}\langle \Phi^2\rangle=2\langle V \Phi\rangle\,,
\\
\label{PV1}
&&\frac{d}{d\tau}\langle V\Phi \rangle=-2\pi\int_{-\infty}^{\infty}P(\pi,V,\tau)V^2dV
+\langle V^2\rangle\,.
\end{eqnarray}
Eq. (\ref{V1}) means that the average square velocity is growing with time, i.e. the soliton undergoes the stochastic acceleration. The law of the
growth of the velocity invariance deviates form the linear, as it would happen for the Brownian diffusion in the momentum space, what happens because
the diffusion coefficient in the Fokker-Planck equation (\ref{FP}) is not a constant, but depends on the angular variable. However, due to the
diffusion one can expect that the phase distribution will tend to   homogeneous, i.e. that ${\cal P}\to 1/(2\pi)$ as $\tau\to\infty$. In this formal
limit one obtains that $\langle V\Phi \rangle\to 0$, $\langle \sin^2\Phi\rangle\to 1/2$ and hence $\langle V^2\rangle\to \tilde{D}\tau $. In other
words, the system (\ref{V1})-(\ref{PV1}), describes random walk which in the limit of large time, approaches the Brownian diffusion in the velocity
space. In that limit the stochastic acceleration, which can be defined as $\widetilde{\gamma}=d\sqrt{\langle V^2\rangle}/d\tau$, would tend to zero
according to the law $\widetilde{\gamma}\propto \tau^{-1/2}$.

\begin{figure}
  \begin{center}
   \begin{tabular}{c}
%        %\scalebox{1.0}[1.0]{\includegraphics[28,586][269,700]{fig1.ps}}
        \scalebox{1.0}[1.0]{\includegraphics{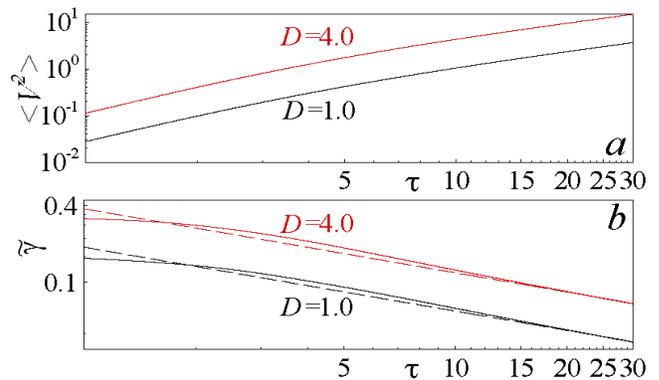}}
   \end{tabular}
   \end{center}
\caption{(Color online) The mean square velocity (panel a) and the stochastic acceleration $\widetilde{\gamma}$  (panel b) of the soliton {\it vs}
time for different values of the dispersion, obtained by numerical integration of Eq. (\ref{FP}) with parameters $L=10.0$ and $k=0.99999$. In panel b
dashed lines depict the approximation of numerical data by the law $\widetilde{\gamma}\propto \tau^{-1/2}$. All axes in panels a,b are represented in
logarithmic scale.} \label{fig2}
\end{figure}

In order to check the above predictions and reveal other features of the stochastic dynamics of a soliton in a ring trap we solved numerically Eq.
(\ref{FP}) subject to the initial condition ${\cal P}(V,\Phi,0)=\langle\delta(\Phi)\delta(V)\rangle$. The results is summarized in Fig.~\ref{fig2}.
In the panel a one observes the predicted monotonic growth of the mean velocity with time, which slightly different from the linear law. In the panel
b one can see that the stochastic acceleration  $\widetilde{\gamma}$ is a monotonically decreasing function, which at sufficiently large times tends
to zero. In particular, at  $\tau\gtrsim 15$ the decreasing of the acceleration with time can be well approximated by the predicted law
$\widetilde{\gamma}\propto \tau^{-1/2}$, as it is shown by dashed curves in the panel b of Fig.~\ref{fig2} (it was verified that in at the same times  $\langle
\sin^2\Phi\rangle \approx 1/2$,  what is in agreement with the analytical predictions).    Also Fig.~\ref{fig2} b shows   that the stochastic acceleration is
larger for larger $D$. The physical explanation of this last fact is simple: the acceleration is generated by the stochastic force, whose intensity
is determined by the dispersion $D$.

\section{Localization of matter induced by the external field}
\label{sec:dyn_loc}

Let us now turn to localized states of a matter in a toroidal trap and concentrate on the states
generated by the constant external electric field, i.e. by $f(\tau)\equiv f_0$. Respectively, we look for stationary solutions of Eq. (\ref{NLS}) in
the form $A=e^{-i\omega \tau}{\cal A}(\zeta)$ and  obtain for ${\cal A}(\zeta)$ the equation:
\begin{eqnarray}
    \label{NLS1}
     -\frac{1}{2} \frac{d^2 {\cal A}}{d \zeta^2}-f_0\cos(\kappa\zeta){\cal  A}+\sigma |{\cal A}|^2{\cal A}=\omega {\cal A}
\end{eqnarray}
which is subject to periodic boundary conditions ${\cal A}(\zeta,\tau)={\cal A}(\zeta+L,\tau)$.

\begin{figure}
  \begin{center}
   \begin{tabular}{c}
        %\scalebox{1.0}[1.0]{\includegraphics[28,586][269,700]{fig1.ps}}
        \scalebox{1.0}[1.0]{\includegraphics{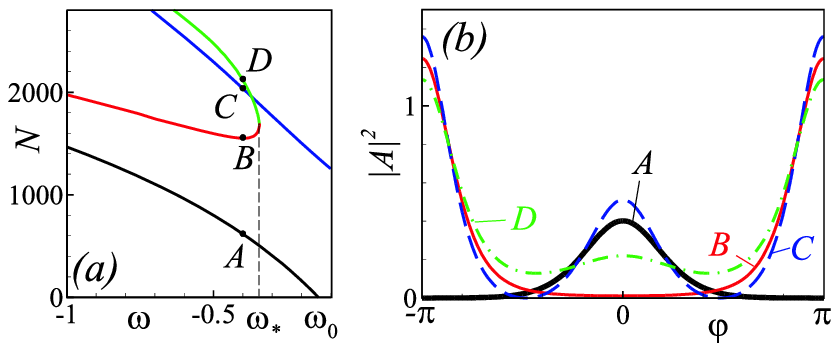}}
   \end{tabular}
   \end{center}
\caption{(Color online) The number of bosons $N$ (for the example of the lithium condensate described in the text) {\textit vs} frequency $\omega$
(panel a) and  shapes of the localized modes at $\omega=-0.4$ (panel b) for the case where $L=10.0$, $f_0=0.3$, $\sigma=-1$.} \label{fig:loc-modes}
\end{figure}

Several lowest branches of the numerically obtained solutions of Eq. (\ref{NLS1}) are shown in  Fig.~\ref{fig:loc-modes}. The lowest branch
approaches zero at the frequency $\omega_0\approx -0.143$ (it is interesting to mention that this frequency coincides with the lowest gap edge of the
spectrum of the Mathieu equation (\ref{NLS1}) considered on the whole axis), where the amplitude of the nonlinear periodic mode is small and it
transforms into the linear periodic Bloch mode at the lowest gap edge. Such a behavior of the branch is similar to that of the strongly localized
modes in  a BECs in the optical lattice~\cite{TS}. The lowest mode A is localized in the vicinity of $\varphi=0$ (Fig.\ref{fig:loc-modes}, b), i.e.
around the minimum of the effective potential and that is why such a mode is stable and can exists even in the linear case, where the two-body
interactions are negligible (here it is important that we are dealing with periodic boundary conditions).  The modes of the upper branches -- B and D
(their typical examples are shown in Fig.\ref{fig:loc-modes} b) -- bifurcate at $\omega_* \approx -0.345$. They are localized  at $\varphi=\pm\pi$,
i.e. near the points where the potential has maxima. As it is clear, this is pure nonlinear effect and occurs due to delicate balance between the
attractive interactions and repulsive forces of the external potential. Such balance can easily be destroyed even by an infinitesimal perturbation,
what allows us to expect instability of the modes. The mode C represents two local maxima of the atomic distributions at $\varphi=0, \pi$. Similar to
the modes B and D we one can expect it to be unstable, what physically can be explained by existence of local atomic maxima at the maxima of the
potential. By direct numerical solution of Eq. (\ref{NLS}) (more specifically by perturbing the mode profiles by the factor $1+0.1\cos{(21\,\zeta)}$
and computing the dynamics until $\tau=1000$) we have verified that, indeed, only the mode A on Fig.~\ref{fig:loc-modes} is dynamically stable, while
the modes B, C, and D are unstable.

\section{Matter soliton as a Kapitza pendulum}
\label{kapitza}

\begin{figure}
  \begin{center}
   \begin{tabular}{c}
        %\scalebox{1.0}[1.0]{\includegraphics[28,586][269,700]{fig1.ps}}
        \scalebox{1.0}[1.0]{\includegraphics{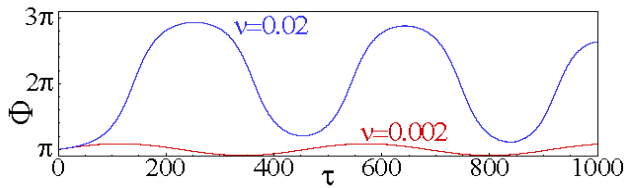}}
   \end{tabular}
   \end{center}
\caption{(Color online) The angular coordinate of the soliton center {\textit vs} time for the soliton motion affected by the rapidly oscillating
external force, obtained numerically from Eq.(\ref{NLS}) with parameters $L=10.0$, $n=0.01$, $\sigma=-1$, $f_0=0.15$, $\Omega=2.0$, and $k=0.99999$.
}
 \label{fig:kap}
\end{figure}

As the final example of nontrivial dynamics of a matter soliton in a toroidal trap we consider dynamical localization induced by a rapidly
oscillating force $f(\tau)=f_0\left[\nu+\cos(\Omega\tau)\right]$. In this case the solitonic motion mimics the famous Kapitza pendulum, which
acquires an additional stable point due rapid oscillation of the pivot~\cite{LL}.  Assuming that the physical conditions of the validity of the
quasi-1D approximation (\ref{NLS}) holds and that the frequency $\Omega$ is large enough, i.e. $\Omega^2\gg \kappa^2 C(k) f_0$, one can perform the
standard analysis (see e.g.~\cite{LL}), i.e. look for a solution of (\ref{NLS}) in a form $X(\tau)+\xi(\tau)$ where $\xi$ is  small, $|\xi|\ll |X|$,
and rapidly varying, and provide averaging over rapid oscillations. Then one arrives at the equation $d^2X/d\tau^2=-\partial U/\partial X$ with the
effective potential
\begin{eqnarray}
    U=-C(k)f_0\left[\nu\cos(\kappa X)+\kappa^2C(k)\frac{f_0}{8\Omega^2}\cos(2\kappa X)\right].
\label{eq:kap_pot}
\end{eqnarray}
If the condition $\kappa^2C(k)f_0/(2\Omega^2\nu)>1$ is met, the effective potential $U$ possesses two stable points: $X=0$ ($\Phi=0$) and $X=L/2$
($\Phi=\pi$). So, it opens the possibility for the new type of soliton moving around the new stable point. Two typical trajectories of the soliton,
obtained by numerical integration of Eq. (\ref{NLS}), are presented in Fig.\ref{fig:kap}. One of the trajectories displays  oscillations around the
new equilibrium point, while the other one shows the large oscillations around the equilibrium point $\Phi=2\pi$ started with the same initial data
but in the case where $\Phi=\pi$ is not an equilibrium any more. The amplitude of large oscillations decay with time because of energy losses of the
soliton in the nonconservative system.

\section{Conclusions}

In the present paper we have shown that dynamics of a matter soliton in a toroidal trap, well reproducing one-dimensional geometry, can be very efficiently governed by
time varying external electric field. In particular, such regimes like dynamical acceleration, stochastic acceleration, localization and
implementation of the Kapitza pendulum can be realized by proper choices of the time dependence of the external force.

Experimental detection of the acceleration can be implemented either by direct imaging of the atomic cloud, which is well localized in space and has well specified trajectory, or by measurement of the atomic distribution in the momentum space displaying shift of the maximum towards higher kinetic energies. Alternatively, one can study the evolution of the atomic cloud  releasing from the trap (by switching the trap off) after some period of accelerating motion. The respective dynamics will be a spreading out cloud whose center of mass is moving with the acquired velocity.

The obtained results were
based on the one-dimensional model, although deduced using the multiple-scale method and thus mathematically controllable. This means that a number
of problem, are still left open. One of them is the limitation on the soliton velocity, and thus acceleration, introduced by lowering the space
dimension, which appears when the solitonic kinetic energy becomes comparable with the transverse kinetic energy. Another limitation on the soliton
acceleration emerges from the fact of  the velocity quantization, when the radius of the ring trap is not large enough. We also left for further
studies the diversity of localized stationary atomic distributions supported by the external filed, indicating only the lowest modes. We thus believe
that the richness of the phenomena which can be observed by simple combination of the trap geometry and varying external field will stimulate new
experimental and theoretical studies.

\acknowledgments

The authors are indebted to H. Michinel for providing additional data on Ref.~\cite{sol_ac}. YVB was supported by the FCT grant SFRH/PD/20292/2004.
VVK was supported by the Secretaria de Stado de Universidades e Investigaci?n (Spain) under the grant SAB2005-0195.   The work was supported by the
FCT (Portugal) and European program FEDER under the grant POCI/FIS/56237/2004.

\end{document}